\documentclass[runningheads]{llncs}
\usepackage{graphicx}
\usepackage{hyperref}
\usepackage{amsmath,amssymb,amsfonts}

\begin{document}

\title{A Comprehensive Survey of Evaluation Techniques
for Recommendation Systems}
\author{Aryan Jadon\inst{1}\orcidID{0000-0002-2991-9913} \and
Avinash Patil\inst{1}\orcidID{0009-0002-6004-370X} }

\authorrunning{Aryan Jadon}
\titlerunning{A Survey of Evaluation Techniques
for Recommendation Systems}

\institute{Juniper Networks, Sunnyvale CA, USA \\
\email{\{aryanj,patila\}@juniper.net}}

\maketitle 

\begin{abstract}

The effectiveness of recommendation systems is pivotal to user engagement and satisfaction in online platforms. As these recommendation systems increasingly influence user choices, their evaluation transcends mere technical performance and becomes central to business success. This paper addresses the multifaceted nature of recommendation system evaluation by introducing a comprehensive suite of metrics, each tailored to capture a distinct aspect of system performance. We discuss 

\begin{itemize}
    \item \textbf{Similarity Metrics}: to quantify the precision of content-based filtering mechanisms and assess the accuracy of collaborative filtering techniques.
    \item \textbf{Candidate Generation Metrics}: to evaluate how effectively the system identifies a broad yet relevant range of items.
    \item \textbf{Predictive Metrics}: to assess the accuracy of forecasted user preferences.
    \item \textbf{Ranking Metrics}: to evaluate the effectiveness of the order in which recommendations are presented.
    \item \textbf{Business Metrics}: to align the performance of the recommendation system with economic objectives.
\end{itemize}

Our approach emphasizes the contextual application of these metrics and their interdependencies. In this paper, we identify the strengths and limitations of current evaluation practices and highlight the nuanced trade-offs that emerge when optimizing recommendation systems across different metrics. 

The paper concludes by proposing a framework for selecting and interpreting these metrics to not only improve system performance but also to advance business goals. This work is to aid researchers and practitioners in critically assessing recommendation systems and fosters the development of more nuanced, effective, and economically viable personalization strategies. Our code is available at \href{https://github.com/aryan-jadon/Evaluation-Metrics-for-Recommendation-Systems}{\textbf{GitHub}}.

\keywords{

Business Metrics,
Candidate Generation Metrics,
Recommendation Systems, 
Predictive Metrics, 
Ranking Metrics,
Similarity Metrics
}

\end{abstract}

\section{Introduction}

Recommendation systems have become an integral component of the digital landscape, influencing the way we discover products, content, and even social connections. From e-commerce to online streaming, these systems underpin user experience by personalizing content and suggesting items that align with individual preferences. The proliferation of such systems has catalyzed a need for robust evaluation methods, as the efficiency of a recommendation system is pivotal to user satisfaction and business success.

While recommendation systems have become increasingly complex and sophisticated, evaluating their performance remains a challenge. Previous research \cite{beel2015comparison} \cite{FKIH20227645}  has focused on specific subsets of performance metrics, often tailored to the domain where the recommendation system is applied\cite{paraschakis2015comparative}. However, a unified approach to performance evaluation, considering the multifaceted aspects of these systems, is largely missing from the literature.

The performance of recommendation systems is multi-dimensional and cannot be encapsulated by a single metric\cite{gunawardana2009survey}. To comprehensively assess these systems, one must consider a variety of metrics, each offering unique insights into different aspects of system performance. This paper introduces \textbf{five key types of metrics} that collectively provide a holistic evaluation framework: \textbf{similarity metrics}, \textbf{candidate generation metrics}, \textbf{predictive metrics}, \textbf{ranking metrics}, and \textbf{business metrics}.

Similarity metrics\cite{li2004similarity} are the cornerstone of content-based and collaborative filtering methods, offering a quantitative measure of how closely items or user preferences align. Candidate generation metrics\cite{hayes2006advancing} ensure a balanced recommendation spectrum, avoiding the pitfalls of overly narrow or excessively broad selections. Predictive metrics\cite{schein2002methods} go a step further, providing an assessment of a system's ability to accurately forecast user ratings or preferences. Ranking metrics \cite{karatzoglou2013learning} are critical when the sequence of recommendations is pivotal, evaluating the order in which items are presented to the user.
Lastly, business metrics\cite{melovic2021strategic} connect system performance with tangible business outcomes, such as sales conversion rates or customer engagement levels, ensuring the recommendation system aligns with overarching business objectives.

In deploying these metrics, one must navigate a landscape rife with trade-offs and complementary relationships, as the improvement in one metric could potentially lead to the detriment of another\cite{jadon2022comprehensive}. Therefore, the selection and interpretation of these metrics must be approached with a nuanced understanding of the recommendation system's goals, context, and the characteristics of the dataset being used.

Through the subsequent sections, this paper will delve into each metric type, elucidating their definitions, applications, and significance in evaluating the efficacy of recommendation systems. By dissecting these metrics, we aim to provide a framework that academics, practitioners, and stakeholders can adopt to gauge the success of their recommendation systems, thereby enabling the continuous advancement of personalized user experiences in the digital domain.

\section{Evaluation Metrics}

\subsection{\textbf{Similarity Metrics}}

Similarity metrics are used to measure the likeness or similarity between items, users, or any relevant entities in a recommendation system. These metrics help in identifying items that are similar to each other, which is crucial for various recommendation techniques like content-based filtering and collaborative filtering. Some of the key Similarity Metrics used in recommendation systems are:

\begin{enumerate}

\item Cosine Similarity
\item Euclidean Distance
\item Jaccard Index
\item Hamming Distance
\item Manhattan Distance
\item Chebyshev Distance
\item Adjusted Cosine Similarity
\item Pearson Correlation Coefficient
\item Spearman Rank Order Correlation Coefficient
\end{enumerate}

\subsubsection{\textbf{Cosine Similarity}} 
Cosine similarity\cite{7577578} is a measure used to determine the similarity between two non-zero vectors in an n-dimensional space, capturing how closely related they are in orientation. It is calculated as the dot product of the vectors divided by the product of their magnitudes. The formula for cosine similarity is:

\[
\text{cos}(\theta) = \frac{\sum_{i=1}^{n} A_i B_i}{\sqrt{\sum_{i=1}^{n} A_i^2} \times \sqrt{\sum_{i=1}^{n} B_i^2}}
\]

This metric ranges from -1 (exactly opposite) to 1 (the same), with 0 typically indicating orthogonality or no similarity. 

\subsubsection{\textbf{Euclidean Distance}}

Euclidean distance\cite{qian2004similarity} is a widely used distance metric that gauges the straight-line distance between two points in Euclidean space. It's calculated by taking the square root of the sum of the squared differences between corresponding elements of the vectors. The formula for the Euclidean distance between two points, \( A \) and \( B \), in an n-dimensional space is:

\[
d(A, B) = \sqrt{\sum_{i=1}^{n} (A_i - B_i)^2}
\]

This metric ranges from 0 to infinity, where 0 indicates that the points are identical.

\subsubsection{\textbf{Jaccard Index}}

The Jaccard Index \cite{real1996probabilistic}, also known as the Jaccard similarity coefficient, is a statistic used for gauging the similarity and diversity of sample sets. It's defined as the size of the intersection divided by the size of the union of two sets. For two sets \( A \) and \( B \), the Jaccard Index \( J \) is given by the formula:

\[
J(A, B) = \frac{|A \cap B|}{|A \cup B|}
\]

Here, \( |A \cap B| \) represents the number of elements common to both sets, while \( |A \cup B| \) is the total number of distinct elements present in either set. The Jaccard Index ranges from 0 to 1, where 0 means there is no overlap between the sets and 1 indicates that the sets are identical. 

\subsubsection{\textbf{Hamming Distance}}

Hamming Distance \cite{norouzi2012hamming} is a metric for comparing two strings of equal length, quantifying the number of positions at which the corresponding symbols are different. It measures the minimum number of substitutions required to change one string into the other, which also corresponds to the minimum number of errors that could have transformed one string into the other. In the context of recommendation systems, it can be particularly useful for comparing user preferences or item characteristics that are represented as binary vectors.

The Hamming Distance, \( d_H \), between two strings (or binary vectors) \( A \) and \( B \) is calculated as:

\[
d_H(A, B) = \sum_{i=1}^{n} \left[ A_i \neq B_i \right]
\]

where \( n \) is the length of the strings, and the notation \( \left[ A_i \neq B_i \right] \) is an indicator function equal to 1 if \( A_i \) and \( B_i \) are different, and 0 if they are the same.

\subsubsection{\textbf{Manhattan Distance}}

Manhattan Distance \cite{strauss2017generalising}, also known as City Block Distance, is a distance metric that measures the sum of the absolute differences between the coordinates of a pair of objects. It is mathematically defined as:

\[
D_{\text{Manhattan}}(A, B) = \sum_{i=1}^{n} |A_i - B_i|
\]

where \( A \) and \( B \) are two vectors in \( n \)-dimensional space, and \( |A_i - B_i| \) denotes the absolute difference between the \( i \)-th components of \( A \) and \( B \).

\subsubsection{\textbf{Chebyshev Distance}}
Chebyshev Distance\cite{gultom2018comparison}, named after Pafnuty Chebyshev, is a distance metric used in multi-dimensional spaces, often in data science and game theory. It represents the maximum difference along any single dimension between two points. For two points \( \mathbf{p} = (p_1, p_2, \ldots, p_n) \) and \( \mathbf{q} = (q_1, q_2, \ldots, q_n) \) in an n-dimensional space, the Chebyshev Distance is defined as:

\[
D_{\text{Chebyshev}}(\mathbf{p}, \mathbf{q}) = \max_{i} \left| p_i - q_i \right|
\]

Here, \( \max_{i} \) signifies taking the maximum of the absolute differences along each dimension. This metric is especially useful in scenarios where the greatest single difference is more significant than the sum of all differences, such as in chess, to calculate the minimum moves for a king, or in clustering and classification tasks in data analytics.

\subsubsection{\textbf{Adjusted Cosine Similarity}}

Adjusted Cosine Similarity \cite{xia2015learning} is a variation of the traditional cosine similarity that accounts for user rating biases. It adjusts the rating vectors for each user by subtracting the user's average rating before applying the cosine similarity formula. This can enhance recommendation system performance by normalizing user ratings and centering them around zero. The formula for Adjusted Cosine Similarity is:

\[
\text{AC}(\theta) = \frac{\sum_{i=1}^{n} (R_{u,i} - \bar{R}_u) \times (R_{v,i} - \bar{R}_v)}{\sqrt{\sum_{i=1}^{n} (R_{u,i} - \bar{R}_u)^2} \times \sqrt{\sum_{i=1}^{n} (R_{v,i} - \bar{R}_v)^2}}
\]

where \( R_{u,i} \) and \( R_{v,i} \) are the ratings given by user \( u \) and user \( v \) to item \( i \), respectively, and \( \bar{R}_u \) and \( \bar{R}_v \) are the average ratings of user \( u \) and user \( v \), respectively. 

\subsubsection{\textbf{Pearson Correlation Coefficient}}

The Pearson Correlation Coefficient (PCC)\cite{sheugh2015note}, denoted as \( r \), is a measure of linear correlation between two sets of data, yielding a value between -1 and 1. A value of 1 implies a perfect positive correlation, -1 a perfect negative correlation, and 0 no correlation at all. For two vectors \( X \) and \( Y \), each with \( n \) elements, PCC is calculated as:

\[
r = \frac{\sum_{i=1}^{n} (X_i - \overline{X})(Y_i - \overline{Y})}{\sqrt{\sum_{i=1}^{n} (X_i - \overline{X})^2} \sqrt{\sum_{i=1}^{n} (Y_i - \overline{Y})^2}}
\]

where \( \overline{X} \) and \( \overline{Y} \) are the means of the \( X \) and \( Y \) vectors, respectively.

In Table 1, we delineate specific scenarios for the application of various Similarity Metrics. This table serves as a concise guide, assisting in selecting the appropriate metric in alignment with the particular characteristics and requirements of each use case. Through this structured presentation, we aim to provide clarity and ease in the decision-making process for choosing the most suitable Similarity Metric, tailored to the needs of distinct scenarios.

\begin{table}
\centering
\caption{Summary of Similarity Metrics.}\label{tab1}
\begin{tabular}{|p{2.5cm}|p{9.5cm}|}
\hline
\textbf{Metric} &  \textbf{Use Cases}\\
\hline
\textbf{Cosine Similarity} & Use cosine similarity when evaluating the orientation, not magnitude, of user or item vectors in recommendation systems, making it ideal for text-based or attribute-rich data where the pattern of interest, rather than the absolute value, is indicative of user preferences. \\ 

\hline
\textbf{Euclidean Distance} & Use Euclidean Distance in evaluating recommendation systems when comparing profiles in a feature space with numerical attributes to gauge similarity based on the 'straight-line' distance between points (user-item pairs). \\

\hline
\textbf{Jaccard Index} & Use the Jaccard Index when evaluating the similarity between users' or items' sets of preferences or choices, particularly when the data is binary and the size of the intersection relative to the union of datasets is of interest. \\

\hline
\textbf{Hamming Distance} & Use Hamming Distance in evaluating recommendation systems when comparing binary vectors of user preferences or item attributes, typically for assessing similarity or diversity in content-based filtering or collaborative filtering with binary datasets. \\

\hline
\textbf{Manhattan Distance} & Use Manhattan Distance in recommendation systems when dealing with high-dimensional, sparse data sets, as it can be more robust to outliers than Euclidean distance and better at capturing differences when multiple dimensions contribute to the dissimilarity. \\
\hline

\textbf{Chebyshev Distance} & Use Chebyshev Distance in recommendation systems when capturing the maximum disparity across dimensions is crucial, especially in high-dimensional spaces where the most significant difference between items is the most informative for recommendations.

\\
\hline
\textbf{Adjusted Cosine Similarity} & Use Adjusted Cosine Similarity when evaluating item-based collaborative filtering systems to account for varying user ratings and reduce bias by subtracting the user's average rating from each of their ratings before computing similarity. \\
\hline
\textbf{Pearson Correlation Coefficient} & Use Pearson Correlation in evaluating recommendation systems when assessing the linear relationship between users’ ratings, especially when the scale of ratings is important and you assume a normal distribution of the underlying data. \\

\hline

\textbf{Spearman Rank Order Correlation Coefficient} & Use Spearman Rank Order Correlation Coefficient in recommendation systems when assessing the strength and direction of a monotonic relationship between ranked variables, particularly useful in scenarios with non-parametric, ordinal data or when evaluating the ranking quality of recommendations.

\\

\hline
\end{tabular}
\end{table}

\subsection{\textbf{Candidate Generation Metrics}}

Candidate Generation Metrics play a pivotal role in the efficacy of recommendation systems, acting as the backbone for filtering and presenting the most relevant options to users. At their core, these metrics are algorithms designed to sift through vast datasets, identifying potential items or services that align closely with a user's preferences, search history, and behavioral patterns. 

This initial step is crucial as it directly influences the quality and relevance of recommendations presented to the user.  By efficiently narrowing down the pool of candidates from potentially millions to a manageable few, these metrics not only enhance the user experience by providing targeted and personalized recommendations but also significantly improve computational efficiency. Furthermore, well-calibrated Candidate Generation Metrics help avoid information overload, ensuring that users are not overwhelmed by too many choices, which can lead to decision paralysis. 

In essence, these metrics are indispensable for creating a tailored, user-centric approach in recommendation systems, leading to increased user engagement, satisfaction, and, ultimately, retention. Some of the key Candidate Generation Metrics used in recommendation systems are :

\begin{enumerate}
\item Novelty
\item Diversity
\item Serendipity
\item Catalog Coverage
\item Distributional Coverage
\end{enumerate}

\subsubsection{\textbf{Novelty}}

Novelty\cite{zhang2013definition}in recommendation systems measures how unexpected the recommended items are to users, focusing on less-known items. Mathematically, for a set of recommended items \( R_u \) to user \( u \), novelty is:

\[ \text{Novelty}(R_u) = \frac{1}{|R_u|} \sum_{i \in R_u} (1 - \text{popularity\_score}(i)) \]

Here, \( \text{popularity\_score}(i) \) is the normalized popularity of item \( i \), calculated as the ratio of users who interacted with \( i \) to the maximum popularity in the catalog. This approach inversely relates the popularity of items to novelty, promoting less popular items to enhance user discovery and exploration. It's an essential metric for ensuring a diverse and engaging recommendation experience.

\subsubsection{\textbf{Diversity}}
Diversity\cite{kunaver2017diversity} in recommendation systems, crucial for maintaining user engagement, is quantified using metrics like Intra-List Diversity (ILD). ILD measures the average dissimilarity between all pairs of items in a recommendation list. Mathematically, for a set of recommended items \( R = \{r_1, r_2, \ldots, r_n\} \), ILD is defined as:

\[
ILD(R) = \frac{2}{n \cdot (n - 1)} \sum_{i=1}^{n-1} \sum_{j=i+1}^{n} dissimilarity(r_i, r_j)
\]

Here, \( n \) is the number of items in \( R \), and \( dissimilarity(r_i, r_j) \) computes the dissimilarity between items \( r_i \) and \( r_j \). ILD ranges from 0 (no diversity) to 1 (maximum diversity), reflecting the variety in recommendations and ensuring that users are exposed to a broad range of options.

\subsubsection{\textbf{Serendipity}}

Serendipity\cite{kotkov2016survey} is a metric used in recommendation systems to quantify the degree to which the recommendations are both unexpected and useful to the user. This concept is crucial in evaluating the effectiveness of a recommendation system, especially in its ability to introduce users to items they might not have discovered otherwise, but find surprisingly relevant and enjoyable.

Mathematically, serendipity can be defined in the context of user-item interactions. Let's consider the following notations:

- \( U \) is the set of users.
- \( I \) is the set of items.
- \( R(u) \) is the set of items recommended to user \( u \).
- \( L(u) \) is the set of items liked by user \( u \).
- \( D(u) \) is the set of items discovered by user \( u \) through the recommendation system.

The serendipity of the system for user \( u \) can be calculated as:

\[
\text{Serendipity}(u) = \frac{|\{i \in R(u) \cap L(u) \cap D(u)\}|}{|R(u)|}
\]

This formula calculates the proportion of recommended items that are both liked and discovered by the user, indicating the element of surprise and relevance in the recommendations.

To get the overall serendipity of the system across all users, we average this value over all users:

\[
\text{Serendipity} = \frac{1}{|U|} \sum_{u \in U} \text{Serendipity}(u)
\]

Here, \( |U| \) denotes the number of users in the system. This overall serendipity score gives an indication of how well the recommendation system performs in terms of introducing relevant yet unexpected items to its users. Higher scores indicate a system's stronger ability to provide serendipitous recommendations.

\subsubsection{\textbf{Catalog Coverage}}

Catalog Coverage\cite{ge2010beyond} is a vital metric in evaluating the breadth of a recommendation system's reach. It measures the proportion of items in the entire catalog that are actually recommended to users, providing insight into the diversity of the system’s suggestions. The formula for Catalog Coverage is:

\[ \text{Catalog Coverage} = \frac{|\text{Unique Items Recommended}|}{|\text{Total Items in Catalog}|} \times 100\% \]

Here, \( |\text{Unique Items Recommended}| \) is the count of distinct items the system has recommended, and \( |\text{Total Items in Catalog}| \) represents the total number of unique items available in the catalog. To compute this, one must tally all unique items recommended over a certain period, count the total items in the catalog, and then divide the former by the latter, expressing the result as a percentage.

This metric is crucial for gauging how well a recommendation system explores and utilizes the full range of available items. A high Catalog Coverage indicates a system that suggests a wide variety of items, potentially appealing to a diverse user base. On the contrary, low Catalog Coverage might indicate a tendency to focus on a limited set of popular items, which could neglect users with unique or niche interests. Therefore, monitoring and optimizing Catalog Coverage is essential for maintaining a balanced and inclusive recommendation system.

\subsubsection{\textbf{Distributional Coverage}}

Distributional coverage\cite{codina2016distributional} is a crucial metric in recommendation systems, focusing on the diversity of recommendations across the entire item catalog. It ensures that a system is not biased towards a few popular items but instead promotes a broader range of choices. To quantify this, distributional coverage is often calculated using entropy, a measure of unpredictability or diversity. For a catalog with \( N \) items, where \( p(i) \) represents the probability of recommending item \( i \), the distributional coverage (\( DC \)) can be expressed as:

\[
DC = -\sum_{i=1}^{N} p(i) \log_2 p(i)
\]

Here, the sum is over all catalog items, and \( p(i) \) is estimated from the frequency of item \( i \)'s appearance in recommendation lists. A higher \( DC \) value indicates a more diverse recommendation pattern, implying a wide array of items being recommended, while a lower value suggests a concentration on fewer items. Balancing this metric with others, such as personalization and relevance, is vital to maintain the effectiveness of the recommendation system while ensuring variety.

Table 2 presents a comprehensive overview of the appropriate application scenarios for various Candidate Generation Metrics. This table serves as a guide, delineating which metrics are most suitable for specific use cases in recommendation systems.

\begin{table}
\centering
\caption{Summary of Candidate Generation Metrics.}\label{tab1}
\begin{tabular}{|p{3cm}|p{9cm}|}
\hline
\textbf{Metric} &  \textbf{Use Cases}\\
\hline
\textbf{Novelty} & Use novelty in evaluating recommendation systems when assessing the system's ability to suggest unexpected, lesser-known, or new items, thereby enhancing user experience by introducing diversity and reducing the filter bubble effect prevalent in personalized recommendations. \\ 
\hline

\textbf{Diversity} & Use Diversity in recommendation systems when the goal is to broaden user exposure beyond familiar items, avoid echo chambers, and enhance user engagement by presenting a varied range of options, thereby catering to a wider spectrum of user interests and preferences. \\ 
\hline

\textbf{Serendipity} & Use serendipity in evaluating recommendation systems when assessing their ability to offer novel, unexpected, yet relevant items, enhancing user experience by introducing diversity beyond typical or predictable suggestions.\\ 

\hline

\textbf{Catalog Coverage} & Use Catalog Coverage to evaluate a recommendation system when assessing its ability to suggest a wide variety of items, particularly important in scenarios where exposing users to a broader selection of the catalog is crucial for enhancing user experience and business objectives.\\ 
\hline

\textbf{Distributional Coverage} & Distributional Coverage is used when evaluating the diversity of recommendations in a system, ensuring a wide range of items are suggested, rather than repeatedly offering popular or similar items, thus enhancing user discovery and experience.\\ 

\hline

\end{tabular}
\end{table}

\subsection{\textbf{Predictive Metrics}}
Predictive metrics are used to assess the predictive accuracy of a recommendation system. These metrics evaluate how well the system predicts user preferences or ratings for items. Some of the key Rating Metrics used in recommendation systems are :

\begin{enumerate}

\item Root Mean Squared Error (RMSE)
\item Mean Absolute Error (MAE)
\item Mean Squared Error (MSE)
\item Mean Absolute Percentage Error (MAPE)
\item $ R^2 $
\item Explained Variance
\end{enumerate}

\subsubsection{\textbf{Root Mean Squared Error (RMSE)}}

Root Mean Squared Error (RMSE)\cite{jahrer2010combining} is a standard way to measure the error of a model in predicting quantitative data. Formally, it represents the square root of the average of the squares of the differences between predicted and observed values. 

In the context of recommendation systems, it quantifies the differences between the ratings predicted by the model and the actual ratings given by the users. The formula for RMSE is given by:

\[
RMSE = \sqrt{\frac{1}{N} \sum_{i=1}^{N} (y_i - \hat{y}_i)^2}
\]

where \( N \) is the number of observations, \( y_i \) is the actual value of an observation, and \( \hat{y}_i \) is the predicted value.

\subsubsection{\textbf{Mean Absolute Error (MAE)}}

Mean Absolute Error (MAE)\cite{cleger2012use} is a metric used to evaluate the accuracy of a prediction model. It calculates the average magnitude of errors in a set of predictions, without considering their direction. The MAE is given by the formula:

\[
\text{MAE} = \frac{1}{n} \sum_{i=1}^{n} |y_i - \hat{y}_i|
\]

where \( n \) is the number of predictions, \( y_i \) is the actual value, and \( \hat{y}_i \) is the predicted value. The absolute difference between the actual and predicted values indicates the error magnitude, and the MAE aggregates these errors across all predictions.

\subsubsection{\textbf{Mean Squared Error (MSE)}}

Mean Squared Error (MSE)\cite{airen2022movie} is a widely used measure of prediction accuracy in recommendation systems, quantifying the difference between predicted and actual values. The MSE is computed by averaging the squares of the errors, i.e., the differences between predicted (\( \hat{y}_i \)) and observed (\( y_i \)) values over \( n \) predictions. The formula for MSE is:

\[
\text{MSE} = \frac{1}{n}\sum_{i=1}^{n} (\hat{y}_i - y_i)^2
\]

The squaring of the errors ensures that larger errors are more prominently reflected in the total, emphasizing the cost of significant deviations and ensuring that the result is always non-negative. 

\subsubsection{\textbf{Mean Absolute Percentage Error (MAPE)}}

Mean Absolute Percentage Error (MAPE)\cite{chicco2021coefficient} is a statistical measure used to assess the accuracy of forecasting models. It represents the average absolute percent error for each data point, omitting the direction of the error. The formula for MAPE is given as:

\[ \text{MAPE} = \left( \frac{100\%}{n} \right) \sum_{i=1}^{n} \left| \frac{y_i - \hat{y}_i}{y_i} \right| \]

Where:
- \( n \) is the number of observations,
- \( y_i \) is the actual value,
- \( \hat{y}_i \) is the forecasted value.

The MAPE is useful because it provides a quick, intuitive percentage error, allowing for comparison across different datasets or models. However, its interpretability can be compromised when dealing with zero or very small actual values.

\subsubsection{\textbf{\( R^2 \)}}

The coefficient of determination, denoted as \( R^2 \), is a crucial metric in evaluating the predictive accuracy of models, including those in recommendation systems. It quantifies the proportion of variance in the dependent variable that is predictable from the independent variables. The formula for \( R^2 \) is:

\[ R^2 = 1 - \frac{SS_{\text{res}}}{SS_{\text{tot}}}\]

where \( SS_{\text{res}} \) is the residual sum of squares (\( \sum (y_i - \hat{y}_i)^2 \)), representing the unexplained variance by the model, and \( SS_{\text{tot}} \) is the total sum of squares (\( \sum (y_i - \bar{y})^2 \)), reflecting the total variance in the observed data. Here, \( y_i \) are the observed values, \( \hat{y}_i \) are the predicted values, and \( \bar{y} \) is the mean of observed data.

\( R^2 \) ranges from 0 to 1, where 0 indicates no explanatory power and 1 indicates perfect prediction. In recommendation systems, a higher \( R^2 \) suggests that the model accurately predicts user ratings or preferences, making it an essential tool for assessing the performance of these systems. It complements other metrics like similarity, classification, and business metrics, offering a comprehensive view of the system's effectiveness and efficiency in personalizing user experiences.

\subsubsection{\textbf{Explained Variance}}

Explained Variance is a key statistical measure in predictive modeling, crucial for evaluating recommendation systems. It quantifies the proportion of variance in the dependent variable (like user ratings) explained by the independent variables in the model. The formula for Explained Variance is:

\[ \text{Explained Variance} = 1 - \frac{\text{Var}(e)}{\text{Var}(Y)} \]

or in a more detailed form:

\[ \text{Explained Variance} = 1 - \frac{\sum_{i=1}^{n} (y_i - \hat{y}_i)^2}{\sum_{i=1}^{n} (y_i - \bar{y})^2} \]

Here, \( \text{Var}(e) \) represents the variance of the model errors, \( \text{Var}(Y) \) is the total variance of the dependent variable, \( y_i \) are the actual values, \( \hat{y}_i \) the predicted values, \( \bar{y} \) the mean of actual values, and \( n \) the number of observations. While closely related to \( R^2 \), the coefficient of determination, Explained Variance offers a nuanced understanding of a model's ability to capture data variability. This metric is vital for researchers in developing efficient recommendation systems as it helps assess the accuracy and reliability of predictions.

In Table 3, we present a comprehensive guide outlining specific scenarios for the application of various Predictive Metrics. This table serves as an essential reference for determining the appropriate metric to employ in distinct contexts, thereby optimizing the effectiveness of predictive analysis.

\begin{table}
\centering
\caption{Summary of Predictive Metrics}\label{tab1}
\begin{tabular}{|p{3cm}|p{9cm}|}
\hline
\textbf{Metric} &  \textbf{Use Cases}\\
\hline
\textbf{Root Mean Squared Error (RMSE)} & Use Root Mean Squared Error (RMSE) to measure the magnitude of prediction errors, giving more weight to larger errors, which is ideal for scenarios where large deviations from actual values are particularly undesirable. \\ 

\hline
\textbf{Mean Absolute Error (MAE))} & Use Mean Absolute Error (MAE) to measure the average magnitude of errors in predictions, prioritizing equal weighting to all deviations regardless of direction, useful for scenarios where all errors are equally important. \\

\hline
\textbf{Mean Squared Error (MSE)} & Use Mean Squared Error (MSE) when evaluating a recommendation system's predictive accuracy for continuous output, as it penalizes larger errors more severely, ensuring the model's robustness by emphasizing the minimization of large prediction errors. \\

\hline
\textbf{Mean Absolute Percentage Error (MAPE)} & Use Mean Absolute Percentage Error (MAPE) to measure forecast accuracy as a percentage, which is useful when you need to understand error magnitude relative to actual values, particularly in inventory or capacity planning where proportional errors are more meaningful. \\

\hline
\textbf{\( R^2 \)} & Use \( R^2 \) (coefficient of determination) in evaluating recommendation systems when assessing the proportion of variance in the observed outcomes (e.g., ratings) that can be predicted from the input variables, providing a measure of how well unseen samples are likely to be predicted.\\

\hline

\textbf{Explained Variance} & Use Explained Variance in evaluating recommendation systems when you aim to measure the proportion of user preference or rating variance that the model successfully captures, indicating the system's overall effectiveness in predicting user behavior accurately.\\

\hline
\end{tabular}
\end{table}

\subsection{\textbf{Ranking Based Metrics}}

Ranking-based metrics are specifically designed to evaluate the quality of the item ranking produced by a recommendation system. These metrics focus on how well the system orders items to maximize user satisfaction. 

\begin{enumerate}

\item Mean Reciprocal Rank (MRR)
\item ARHR@k
\item Normalized Discounted Cumulative Gain (nDCG@K)
\item Precision@k
\item Recall@k
\item F1@K
\item Average Recall@k
\item Average Precision@k
\item MAP

\end{enumerate}

\subsubsection{\textbf{Mean Reciprocal Rank (MRR)}}

Mean Reciprocal Rank (MRR)\cite{shi2012climf} is a statistical measure used to evaluate the performance of recommendation systems or information retrieval systems, focusing on the rank of the first correct answer. For a set of queries, MRR is the average of the reciprocal ranks of results for the queries. The reciprocal rank of a query response is the inverse of the rank of the first correct item. If the correct item is at rank \( k \), the reciprocal rank is \( \frac{1}{k} \). For a set of \( Q \) queries, MRR is calculated as:

\[
MRR = \frac{1}{|Q|} \sum_{i=1}^{|Q|} \frac{1}{\text{rank}_i}
\]

where \( |Q| \) is the number of queries and \( \text{rank}_i \) is the position of the first relevant item for the \( i \)-th query.

\subsubsection{\textbf{Average Reciprocal Hit-Rank at K (ARHR@k)}}
The Average Reciprocal Hit-Rank at K (ARHR@k)\cite{elkorany2013semantic} is a crucial metric in evaluating recommendation systems, particularly emphasizing the ranking efficiency of the first relevant recommendation within the top-K items. It's defined by the formula:

\[
\text{ARHR@k} = \frac{1}{|U|} \sum_{u \in U} \sum_{i=1}^{k} \frac{\delta(i, u)}{i}
\]

Here, \(|U|\) represents the total number of users, \(k\) is the cut-off rank for top recommendations, and \(\delta(i, u)\) is an indicator function that equals 1 if the item at rank \(i\) is relevant to user \(u\), and 0 otherwise. This metric calculates the average of the reciprocal ranks of the first 'hit' or relevant item for each user but only if it appears within the top-K suggestions. A higher ARHR@k score indicates that the system not only accurately identifies relevant items but also ranks them highly, enhancing user experience and satisfaction. This metric is particularly valuable in scenarios where the prominence of recommendations significantly impacts user engagement.

\subsubsection{\textbf{Normalized Discounted Cumulative Gain (nDCG)}}

Normalized Discounted Cumulative Gain (nDCG)\cite{busa2012apple} is a measure of ranking quality that captures the performance of ranking algorithms in recommendation systems. It evaluates how well the predicted ranking of items corresponds to the ideal ranking, considering the relevance of each item. The DCG is computed as:

\[
\text{DCG}_p = \sum_{i=1}^{p} \frac{2^{rel_i} - 1}{\log_2(i+1)}
\]

where \( rel_i \) is the relevance score of the item at position \( i \) and \( p \) is the number of ranked items. nDCG is obtained by normalizing DCG with the ideal DCG (iDCG), which is the DCG score obtained by the perfect ranking:

\[
\text{nDCG}_p = \frac{\text{DCG}_p}{\text{iDCG}_p}
\]

A perfect ranking would result in an nDCG of 1, while any deviation from the ideal ranking results in an nDCG less than 1.

\subsubsection{\textbf{Precision@k}}

Precision@k is a performance metric that evaluates the relevance of a list of recommended items. It measures the proportion of recommended items in the top-k set that are relevant to the user. The formula for Precision@k is given by:

\[
\text{Precision@k} = \frac{\text{Number of relevant items in top-k}}{k}
\]

Here, "relevant items" are those that are deemed to be of interest to the user based on some ground truth, such as past user behavior or explicit ratings. The metric provides a straightforward indication of recommendation quality at a fixed list size \( k \). 

\subsubsection{\textbf{Recall@k}}

Recall@k is a metric used to evaluate recommendation systems based on how many relevant items are selected out of all possible relevant items. Specifically, for a set of queries, it measures the proportion of relevant items found in the top-k recommendations. The formula for Recall@k is given by:

\[
\text{Recall@k} = \frac{|\{\text{Relevant items}\} \cap \{\text{Top-k recommended items}\}|}{|\{\text{Relevant items}\}|}
\]

Here, \( |\{\text{Relevant items}\}| \) is the number of relevant items, and \( |\{\text{Relevant items}\} \cap \{\text{Top-k recommended items}\}| \) is the number of relevant items that are in the top-k recommendations.

\subsubsection{\textbf{F1@K (F1 score at K)}}
The F1 score is the harmonic mean of precision and recall. At a specific cut-off point K, it is calculated as:
\[
F1@K = 2 \times \frac{\text{Precision@K} \times \text{Recall@K}}{\text{Precision@K} + \text{Recall@K}}
\]
where Precision@K and Recall@K are the precision and recall calculated at the cut-off K.

\subsubsection{\textbf{Average Recall@K}}

Recall@K for a single user is the proportion of relevant items that are in the top K recommendations. The Average Recall@K across all users is:
\[
\text{Average Recall@K} = \frac{1}{U} \sum_{u=1}^{U} \frac{|\text{Relevant Items}_u \cap \text{Recommended Items}_u@K|}{|\text{Relevant Items}_u|}
\]

where U is the total number of users, Relevant Items u is the set of relevant items for user u, and Recommended Items u@K is the set of top K recommended items for user u.

\subsubsection{\textbf{Average Precision@K}}
Precision@K for a single user is the proportion of recommended items in the top K that are relevant. The Average Precision@K is:
\[
\text{Average Precision@K} = \frac{1}{U} \sum_{u=1}^{U} \frac{|\text{Relevant Items}_u \cap \text{Recommended Items}_u@K|}{K}
\]

\subsubsection{\textbf{Mean Average Precision (MAP)}}
MAP considers the order of recommendations. It is the mean of the Average Precision at each point a relevant item is retrieved, averaged over all users:
\[
\text{MAP} = \frac{1}{U} \sum_{u=1}^{U} \left( \frac{1}{|\text{Relevant Items}_u|} \sum_{k=1}^{|\text{Recommended Items}_u|} \text{Precision@k} \times \text{rel}_u(k) \right)
\]
where rel u(k) is an indicator function that is 1 if the item at rank k is relevant to user u and 0 otherwise.

Table 4 enumerates the specific scenarios for applying Ranking Metrics.

\begin{table}
\centering
\caption{Summary of Ranking Metrics}\label{tab1}
\begin{tabular}{|p{3cm}|p{9cm}|}
\hline
\textbf{Metric} &  \textbf{Use Cases}\\

\hline
\textbf{Mean Reciprocal Rank (MRR)} & 
Use Mean Reciprocal Rank (MRR) when evaluating the effectiveness of a recommendation system at returning the first relevant item in a ranked list of results, emphasizing the importance of the top-most recommendation. \\

\hline

\textbf{ARHR@k (Average Reciprocal Hit Rank at k)} & Use ARHR@k (Average Reciprocal Hit Rank at k) to measure the average quality of recommendations by considering the reciprocal of the rank at which relevant items are found in the top k positions. It's useful for scenarios where ranking accuracy and the order of recommendations are crucial factors.\\

\hline

\textbf{Normalized Discounted Cumulative Gain (nDCG)} & Use nDCG when ranking relevance matters, and you need to evaluate the quality of the recommendations in a list, considering the position of highly relevant items. \\ 

\hline

\textbf{Mean Reciprocal Rank (MRR)} & 
Use Mean Reciprocal Rank (MRR) when evaluating the effectiveness of a recommendation system at returning the first relevant item in a ranked list of results, emphasizing the importance of the top-most recommendation. \\

\hline
\textbf{Precision@k} & Use Precision@k in recommendation systems when evaluating the proportion of relevant items among the top-k recommendations, particularly when the cost of false positives is high and the focus is on the quality of the top few recommendations. \\

\hline
\textbf{Recall@k} & Use Recall@k in recommendation systems when prioritizing the model's ability to capture all relevant items in the top-k suggestions is more important than avoiding irrelevant ones, especially when the cost of missing a relevant recommendation is high. \\

\hline
\textbf{F1@K} & Use F1@K (F1 score at K) for scenarios where the focus is on the precision of the top K recommendations, such as content curation or ranking tasks, to ensure that the most relevant items are presented to users within a limited set. \\

\hline
\textbf{Average Recall@K} & Use Average Recall@K when you need to assess the proportion of relevant items that a recommendation system successfully retrieves within the top K recommendations, prioritizing comprehensive coverage of relevant content over precision. \\

\hline
\textbf{Average Precision@K} & Use Average Precision@k when evaluating the effectiveness of a recommendation system for scenarios where the order of recommended items matters, such as search engine result rankings or content playlists, as it quantifies the precision of the top k recommendations. \\

\hline
\textbf{Mean Average Precision (MAP)} & Use Mean Average Precision (MAP) when evaluating the quality of a ranked list of items in recommendation systems particularly when relevance varies across items, as it provides a comprehensive measure of precision at different points in the ranking, emphasizing the importance of higher-ranked items.\\

\hline

\end{tabular}
\end{table}

\subsection{\textbf{Business Metrics}}

Business metrics play a crucial role in assessing the performance and impact of recommendation systems (RS) in various domains\cite{dijkman2011similarity}. One essential metric is \textbf{Click-through Rate (CTR)}, which measures the number of clicks generated by recommendations. Higher CTR indicates that recommendations are more relevant, making it a popular metric in the news recommendation domain. Platforms like Google News and Forbes utilize CTR to gauge the effectiveness of their recommendations. Personalized suggestions based on CTR have been shown to increase clicks by up to 38\% compared to popularity-based systems, underscoring its importance.

\textbf{Adoption and conversion} metrics provide deeper insights into user behavior. While CTR indicates clicks, it doesn't determine if those clicks led to conversions or purchases. Platforms like YouTube and Netflix use alternative adoption measures such as "Long CTR" (counting clicks when users watch a specific percentage of a video) and "Take rate" (counting views after recommendations) to assess user engagement and conversion. In cases where an item cannot be viewed directly, domain-specific measures, such as the number of contacts made with an employer after a job offer recommendation on LinkedIn, become crucial.

\textbf{Sales and revenue} metrics are ultimately what matter for businesses. Although CTR and adoption metrics are informative, changes in sales and revenue reflect the actual impact on the bottom line. However, attributing improvements solely to RS can be challenging, as users may have made purchases anyway, making it necessary to consider the broader business context.

\textbf{Measuring the effects on sales distribution} is another critical aspect. This metric directly compares sales before and after the introduction of RS. However, it requires an understanding of how the shifts in sales distribution impact diversity at the individual level. Efforts to maintain diversity may be needed to prevent unintended consequences.

\textbf{User behavior and engagement} metrics highlight the impact of RS on user activity and retention. Recommendations often increase user engagement, and a positive correlation between customer engagement and retention is observed in various domains, such as Spotify. However, measuring this can be challenging, especially when churn rates are low. These metrics collectively provide a comprehensive view of RS performance, helping businesses make data-driven decisions to enhance recommendation systems.

\section{Experiments and Results}

In this study, we conducted experiments on three different MovieLens datasets, namely \textbf{MovieLens 100k, MovieLens 1m, and MovieLens 10m}, to evaluate the performance of our recommendation system. We aimed to assess various metrics to gain insights into the quality and effectiveness of our recommendation algorithms. The results of these experiments are summarized in Tables 5 and 6 below.

We present the experimental results obtained from the analysis of two datasets: \textbf{Amazon Electronics Dataset} and \textbf{Amazon Movies and TV Dataset}. The experiments were conducted to evaluate the performance of various similarity metrics in the context of recommendation systems. The results are summarized in the following table:

\begin{table}[htbp]
\centering
\caption{Similarity Metrics Experiments on Amazon Electronics Dataset}
\begin{tabular}{|c|c|c|c|}
\hline
\textbf{Similarity Measure} & \textbf{EPOC} & \textbf{Loss} & \textbf{AUC} \\
\hline

\textbf{Adjusted Cosine Similarity} & 4 & 0.7289 & 0.9531 \\
\hline

\textbf{Chebyshev Distance Similarity} & 4 & 0.7105 & 0.2423 \\
\hline

\textbf{Cosine Similarity} & 4 & 0.7128 & \textbf{0.9531} \\
\hline

\textbf{Euclidean Distance Similarity} & 4 & 0.7107 & 0.1898 \\
\hline

\textbf{Hamming Distance} & 4 & 0.7054 & 0.2740 \\
\hline

\textbf{Jaccard Index Similarity} & 4 & 0.7167 & 0.9060 \\
\hline

\textbf{Manhattan Distance Similarity} & 4 & 0.7098 & 0.1849 \\
\hline

\textbf{Pearson Correlation Coefficient Similarity} & 4 & 0.7130 & 0.9158 \\
\hline

\textbf{Spearman Rank Order Correlation Coefficient Similarity} & 4 & 0.7111 & -0.5373 \\
\hline

\end{tabular}
\end{table}

These results provide valuable insights into the performance of different similarity measures when applied to recommendation systems, with AUC serving as a key metric for evaluating the quality of recommendations. The experiments were conducted with a consistent number of epochs and batch size (2800) across all similarity measures for a fair comparison.

\begin{table}[htbp]
\centering
\caption{Similarity Metrics Experiments on Amazon Movies and TV Dataset}
\begin{tabular}{|c|c|c|c|}
\hline
\textbf{Similarity Measure} & \textbf{EPOC} & \textbf{Loss} & \textbf{AUC} \\
\hline

\textbf{Adjusted Cosine Similarity} & 4 & 0.6177 & 0.9444 \\
\hline

\textbf{Chebyshev Distance Similarity} & 4 & 0.6051 & 0.2364 \\
\hline

\textbf{Cosine Similarity} & 4 & 0.6123 & \textbf{0.9449} \\
\hline

\textbf{Euclidean Distance Similarity} & 4 & 0.6044 & 0.1738 \\
\hline

\textbf{Hamming Distance} & 4 & 0.6047 & 0.2507 \\
\hline

\textbf{Jaccard Index Similarity} & 4 & 0.6040 & 0.8765 \\
\hline

\textbf{Manhattan Distance Similarity} & 4 & 0.6046 & 0.1690 \\
\hline

\textbf{Pearson Correlation Coefficient Similarity} & 4 & 0.6088 & 0.9181 \\
\hline

\textbf{Spearman Rank Order Correlation Coefficient Similarity} & 4 & 0.6049 & -0.5654 \\
\hline

\end{tabular}
\end{table}

Table 7 presents the catalog coverage, distributional coverage, novelty, diversity, and serendipity metrics for each dataset. These metrics provide valuable insights into the recommendation system's ability to cover a wide range of items, recommend items not previously seen by users, introduce novel items, maintain diversity, and offer serendipitous recommendations. The results indicate that as the dataset size increases, the recommendation system's performance in terms of these metrics generally improves.

\begin{table}[h]
\centering
\caption{Candidate Generation Metrics on MovieLens Datasets}
\begin{tabular}{|l|c|c|c|c|c|}
\hline
\textbf{Dataset} & \textbf{Catalog Coverage} & \textbf{Distributional Coverage} & \textbf{Novelty} & \textbf{Diversity} & \textbf{Serendipity} \\
\hline
\textbf{100k} & 0.3951 & 8.2069 & 11.5816 & 0.8801 & 0.8750 \\
\hline
\textbf{1m}   & 0.2438 & 7.6453 & 15.2664 & 0.9594 & 0.9365 \\
\hline
\textbf{10m}  & 0.1570 & 7.4687 & 20.0912 & 0.9708 & 0.9809 \\
\hline
\end{tabular}
\label{table:results}
\end{table}

We also evaluated two collaborative filtering algorithms, ALS (Alternating Least Squares) and SVD (Singular Value Decomposition), on each dataset using a fixed value of K (number of recommendations). The results of these experiments are summarized in the following table:

Table 8 provides an overview of the performance metrics for each combination of dataset and algorithm, including training time, predicting time, Root Mean Square Error (RMSE), Mean Absolute Error (MAE), Mean Squared Error (MSE), Mean Absolute Percentage Error (MAPE), R-squared (R2), and Explained Variance. These metrics allow us to assess the effectiveness of ALS and SVD in providing recommendations on different scales of MovieLens datasets.

\begin{table}[h]
\centering
\caption{Predictive Metrics on MovieLens Datasets}
\begin{tabular}{|c|c|c|c|c|c|c|}
\hline
\textbf{Dataset} & \textbf{100k} & \textbf{100k} & \textbf{1m} & \textbf{1m} & \textbf{10m} & \textbf{10m} \\
\hline
\textbf{Algorithm} & \textbf{ALS} & \textbf{SVD} & \textbf{ALS} & \textbf{SVD} & \textbf{ALS} & \textbf{SVD} \\
\hline
\textbf{K} & 10 & 10 & 10 & 10 & 10 & 10 \\
\hline
\textbf{Train Time (s)} & 5.1556 & 0.5020 & 3.7938 & 5.7888 & 19.0998 & 56.7193 \\
\hline
\textbf{Predicting Time (s)} & 0.0442 & 0.1992 & 0.0165 & 0.7162 & 0.0168 & 8.9013 \\
\hline
\textbf{RMSE} & 0.9659 & 0.9425 & 0.8583 & 0.8841 & 0.7984 & 0.8077 \\
\hline
\textbf{MAE} & 0.7521 & 0.7448 & 0.6761 & 0.6961 & 0.6184 & 0.6211 \\
\hline
\textbf{MSE} & 0.9329 & 0.8883 & 0.7367 & 0.7817 & 0.6375 & 0.6524 \\
\hline
\textbf{MAPE} & 28.8383 & 29.9627 & 25.2899 & 27.0609 & 24.8207 & 25.8009 \\
\hline
\textbf{R2} & 0.2685 & 0.2999 & 0.4159 & 0.3761 & 0.4393 & 0.4201 \\
\hline
\textbf{Explained Variance} & 0.2648 & 0.2999 & 0.4120 & 0.3762 & 0.4333 & 0.4201 \\
\hline
\end{tabular}
\end{table}

We evaluated the performance of seven different recommendation algorithms, namely ALS\cite{pilaszy2010fast}, SAR \cite{gong2010collaborative}, SVD \cite{graham2019microsoft}, NCF\cite{sang2021knowledge}, BPR \cite{rendle2012bpr}, BiVAE \cite{truong2021bilateral}, and LightGCN\cite{he2020lightgcn}, across both datasets. The following table summarizes the key metrics for each algorithm on the MovieLens 100k dataset.

Table 9 provides a comprehensive overview of the algorithm performance in terms of various evaluation metrics, including Mean Average Precision (MAP), normalized Discounted Cumulative Gain (nDCG@k), Precision@k, Recall@k, F1@k, Mean Reciprocal Rank (MRR), Average Rank-based Half-life Reciprocal (ARHR@k), Average Recall@k, and Average Precision@k. These metrics offer valuable insights into the recommendation quality and efficiency of each algorithm on the MovieLens 100k dataset.

\begin{table}[htbp]
\centering
\caption{Ranking Metrics on MovieLens 100k Dataset}
\begin{tabular}{|c|c|c|c|c|c|c|c|c|}
\hline
\textbf{Dataset} & \textbf{100k} & \textbf{100k} & \textbf{100k} & \textbf{100k} & \textbf{100k} & \textbf{100k} & \textbf{100k} \\
\hline
\textbf{Algorithm} & \textbf{ALS} & \textbf{SAR} & \textbf{SVD} & \textbf{NCF} & \textbf{BPR} & \textbf{BiVAE} & \textbf{LightGCN} \\
\hline
\textbf{K} & 10 & 10 & 10 & 10 & 10 & 10 & 10 \\
\hline
\textbf{Train time (s)} & 4.7152 & 0.2113 & 0.3912 & 15.9303 & 3.5350 & 35.8652 & 20.4386 \\
\hline
\textbf{Recommending time (s)} & 0.0744 & 0.1555 & 6.2471 & 3.4046 & 0.6358 & 0.6011 & 0.0313 \\
\hline
\textbf{MAP} & 0.0035 & 0.1140 & 0.0121 & 0.0945 & 0.1340 & 0.1467 & 0.1227 \\
\hline
\textbf{nDCG@k} & 0.0342 & 0.3938 & 0.0944 & 0.3687 & 0.4450 & 0.4753 & 0.4216 \\
\hline
\textbf{Precision@k} & 0.0398 & 0.3406 & 0.0891 & 0.3288 & 0.3887 & 0.4152 & 0.3644 \\
\hline
\textbf{Recall@k} & 0.0144 & 0.1854 & 0.0301 & 0.1640 & 0.2166 & 0.2254 & 0.1980 \\
\hline
\textbf{F1@k} & 0.0212 & 0.2401 & 0.0450 & 0.2189 & 0.2781 & 0.2921 & 0.2565 \\
\hline
\textbf{MRR} & - & 0.7180 & 0.4487 & 0.6649 & 0.7273 & 0.7644 & 0.7208 \\
\hline
\textbf{ARHR@k} & - & 0.3466 & 0.3183 & 0.3355 & 0.3401 & 0.3412 & 0.3474 \\
\hline
\textbf{Average Recall@k} & - & 0.2958 & 0.0936 & 0.3082 & 0.3144 & 0.3505 & 0.2903 \\
\hline
\textbf{Average Precision@k} & - & 0.4144 & 0.1510 & 0.3868 & 0.4428 & 0.4854 & 0.4343 \\
\hline
\end{tabular}
\end{table}

Table 10 provides a comprehensive overview of the algorithm performance on the MovieLens 1M dataset.

\begin{table}[htbp]
\centering
\caption{Ranking Metrics on MovieLens 1M Dataset}
\begin{tabular}{|c|c|c|c|c|c|c|c|}
\hline
\textbf{Dataset} & \textbf{1M} & \textbf{1M} & \textbf{1M} & \textbf{1M} & \textbf{1M} & \textbf{1M} & \textbf{1M} \\
\hline
\textbf{Algorithm} & \textbf{ALS} & \textbf{SAR} & \textbf{SVD} & \textbf{BPR} & \textbf{BiVAE} & \textbf{LightGCN} & \textbf{NCF} \\
\hline
\textbf{K} & 10 & 10 & 10 & 10 & 10 & 10 & 10 \\
\hline
\textbf{Train time (s)} & 11.3532 & 1.7349 & 4.7138 & 50.0558 & 569.2301 & 1958.8973 & 331.4241 \\
\hline
\textbf{Recommending time (s)} & 0.1666 & 3.2854 & 90.2653 & 11.9814 & 28.6078 & 0.6001 & 54.6452 \\
\hline
\textbf{MAP} & 0.0022 & 0.0650 & 0.0088 & 0.0838 & 0.0898 & 0.0889 & 0.0667 \\
\hline
\textbf{nDCG@k} & 0.0283 & 0.3121 & 0.0902 & 0.3931 & 0.4247 & 0.4241 & 0.3630 \\
\hline
\textbf{Precision@k} & 0.0358 & 0.2803 & 0.0841 & 0.3603 & 0.3890 & 0.3864 & 0.3337 \\
\hline
\textbf{Recall@k} & 0.0105 & 0.1106 & 0.0224 & 0.1415 & 0.1465 & 0.1468 & 0.1144 \\
\hline
\textbf{F1@k} & 0.0163 & 0.1587 & 0.0354 & 0.2032 & 0.2129 & 0.2128 & 0.1704 \\
\hline
\textbf{MRR} & - & 0.6407 & 0.4512 & 0.6776 & 0.7074 & 0.7092 & 0.6729 \\
\hline
\textbf{ARHR@k} & - & 0.3405 & 0.3276 & 0.3292 & 0.3305 & 0.3334 & 0.3320 \\
\hline
\textbf{Average Recall@k} & - & 0.1842 & 0.0714 & 0.2223 & 0.2474 & 0.2432 & 0.2083 \\
\hline
\textbf{Average Precision@k} & - & 0.3418 & 0.1455 & 0.4009 & 0.4434 & 0.4460 & 0.3983 \\
\hline
\end{tabular}
\end{table}

\section{Conclusion}

Evaluating the efficacy of recommendation systems in business contexts is multifaceted, requiring more than just accuracy-related metrics. Although rank-aware accuracy metrics serve as a good starting point for generating candidate items via machine learning techniques, they often overlook crucial aspects like diversity and novelty, which are key to customer satisfaction. The experiments conducted on various datasets underscore the limitations of accuracy-focused assessments. 

Importantly, real-world business value and user satisfaction emerge not solely from these ML metrics but significantly from user feedback. This is evident in the enhancements observed in click-through rates, sales, revenue, and related derivatives when A/B testing is incorporated. Consequently, A/B testing emerges as an essential tool, bridging the gap between ML model predictions and tangible business outcomes. By aligning business strategies with machine learning models through comprehensive evaluation, including user feedback, recommendation systems can be fine-tuned to achieve harmony between technical performance and business success.

\bibliographystyle{splncs04}
\nocite{*}
\bibliography{references}
\end{document}